\documentclass[aps,prb,
twocolumn, 
               a4paper,amsmath,amssymb,superscriptaddress]{revtex4-1}

\usepackage[utf8]{inputenc}
\usepackage[english]{babel}
\usepackage{graphicx}
\usepackage{feynmp}
\usepackage{bm}
\usepackage{color}

\usepackage{color}
\definecolor{Green}{rgb}{0,0.7,0}

\newcommand{\e}{ {\rm e}}
\newcommand{\ET}{$\alpha$-(BEDT-TTF)$_2$I$_3$}
\newcommand{\bk}{\bm{k}}

\newcommand{\bkD}{\bm{k}_{\rm D}}

\newcommand{\ep}{\epsilon }

\newcommand{\kx}{k_x}
\newcommand{\ky}{k_y}

\newcommand{\g}{\gamma }
\newcommand{\bq}{\bm{q}}
\newcommand{\BETS}{$\alpha$-(BETS)$_2$I$_3$}
\newcommand{\br}{\bm{r}}
\newcommand{\bG}{\bm{G}}

\newcommand{\muz}{\mu_0}

\newcommand{\tep}{\tilde{\epsilon}}

\newcommand{\suzu}[1]{\textcolor{black}{#1}}

\begin{document}

\title{
Spin-Orbit  Coupling Effect on the Seebeck Coefficient 
 in Dirac Electron Systems in $\alpha$-(BETS)$_2$I$_3$
}


\date
{Received  2 September 2025; accepted 10 October 2025; published 10 November 2025}

\author{Yoshikazu \surname{Suzumura}}
\email{suzumura.yoshikazu.k7@a.mail.nagoya-u.ac.jp} 
 \affiliation{Department of Physics, Nagoya University,
             Nagoya 464-8602, Japan}            
\author{Takao \surname{Tsumuraya}}
\email{tsumu@kumamoto-u.ac.jp} 
 \affiliation{Magnesium Research Center, Kumamoto University, Kumamoto 860-8555, Japan} 
\author{Masao \surname{Ogata}}
\email{ogata@phys.s.u-tokyo.ac.jp} 
 \affiliation{Department of Physics, University of Tokyo, Bunkyo, Tokyo 113-0033, Japan \\ and
 National Institute of Advanced Industrial Science and Technology, Tsukuba, Ibaraki 305-8568, Japan}

\begin{abstract}
The Seebeck coefficient, $S=L_{12}/(TL_{11})$, which is 
 proportional to a ratio of the thermoelectric conductivity $L_{12}$ to the 
electric conductivity $L_{11}$  with $T$ being temperature is examined  
 for two-dimensional Dirac electrons in  the three-quarter filled  organic conductor, $\alpha$-(BETS)$_2$I$_3$, 
 [BETS = BEDT-TSeF = bis(ethylenedithio)tetraselenafulvalene] 
at ambient pressure. 
 Using a tight-binding model obtained with the  first-principles 
relativistic density-functional theory method [Tsumuraya and Suzumura, Eur. Phys. J. B {\bf 94}, 17 (2021)], 
  we calculate  $S$ 
in the presence of  
     the impurity  and electron--phonon scatterings.   
 We show  that 
 $S_x < 0$ and $S_y >0$ at high temperatures, where 
 $S_x$ ($S_y$)  denotes $S$ perpendicular (parallel) to the molecular stacking  axis.   
 There is a sign change of $S_y$  with increasing  $T$. 
 We find that at low temperatures the absolute value of $S$ is enhanced by the spin-orbit coupling. 
 The Seebeck coefficient is examined by dividing it  into components of 
  the conduction and valence bands;  
  we find that  the electron and hole contributions compete with  each other. 
 Such $T$ dependence of $S$  is clarified  using  
  the spectral conductivity,  
 which determines $L_{12}$ and $L_{11}$.\\

DOI: 10.1103/9htn-m1pp
\end{abstract}


\maketitle


\section{Introduction}
 Two-dimensional massless Dirac fermions (MDFs) 
  with a linear spectrum around a Dirac point 
    have been studied extensively.~\cite{Novoselov2005_Nature438}  
  Among them,  bulk material has been found in 
  the  organic conductor,~\cite{Kajita_JPSJ2014}  
 \ET, [BEDT-TTF =   
bis(ethylenedithio)tetrathiafulvalene]~\cite{Mori1984}.
 This material exhibits    a zero-gap state  
   in the  energy band due to  
 the Dirac cone~\cite{Kobayashi2004,Katayama2006_JPSJ75},   
 which is  calculated   using  a tight-binding (TB) model with   
   transfer energies  estimated from 
 the extended H\"uckel method~\cite{Kondo2005,Kondo2009}. 
   The transport properties of such a system are affected by  
     the density of states (DOS), which  
      reduces linearly to zero at the Fermi energy~\cite{Kobayashi2004}.  
 A first-principles calculation based on density functional theory (DFT) 
 confirmed the existence of MDFs in \ET~at ambient pressure~\cite{Kino2006}. 
 Under hydrostatic pressure,  a DFT band structure~\cite{Alemany} 
  was calculated using the crystal structure 
  determined by x-ray diffraction measurements at 1.76GPa~\cite{Kondo2009}.
Despite these investigations, the impact of pressure on the transfer energies calculated using first-principles methods has not been explored.
 Thus,  a  different model  was presented  by  deriving  
 an $ab$ $initio$ TB model~\cite{Suzumura_JPSJ_2024}
 with  the experimentally obtained crystal structure~\cite{Kondo2009}. 

Characteristic properties  of  the MDFs appear in  the temperature $T$  
 dependence of various physical quantities. The $T$-linear behavior of  
 the magnetic susceptibility~\cite{TakanoJPSJ2010,Hirata2016}
 and the sign change of the Hall coefficient,\cite{Tajima_Hall2012} 
 are  understood with theories 
 using the four-band model~\cite{Katayama_EPJ} and 
  the two-band model~\cite{Kobayashi2008,Montambaux_2008}, respectively. 
However, almost-$T$-independent 
conductivity~\cite{Kajita1992,Tajima2000,Tajima2002,Tajima2007,Liu2016} 
cannot be understood with a model 
 with only impurity scattering~\cite{Neto2006}.
If we consider the effect of electron-phonon (e--p) interaction, 
a nearly-constant conductivity can be understood using a simple two-band 
model of the Dirac cone without tilting~\cite{Suzumura_PRB_2018}.   
 Because  the energy band in the actual organic conductor   
 deviates  from the linear spectrum~\cite{Katayama2006_cond},   
  the conductivity was calculated using the TB model   
 of \ET  \;  with the e--p interaction~\cite{Suzumura_JPSJ2021}. 

In organic conductors,  Dirac electrons are 
 also expected for  \BETS \;
[BETS = BEDT-TSeF =bis(ethylenedithio)tetraselenafulvalene] 
 since the almost constant resistivity  has been  observed  
  at high temperatures~\cite{Inokuchi1995}.  
The gradual increase in  resistivity at low temperatures in \BETS \; 
 shows an  insulating behavior but no symmetry breaking,  
 in contrast to the charge ordering  of 
 \ET.~\cite{Takahashi2006,Hiraki,Kitou2021}. 
Taking account of the  spin-orbit interaction (SOI)
 in organic conductors~\cite{Winter},  
  an explicit  TB model with SOI 
 was derived  for \ET, whose 
  transfer energies are complex and provides  the 
 Berry curvature of spin subbands~\cite{Osada}.  
As for \BETS,
  the band calculation based on the extended H\"uckel method~\cite{Kondo2009} 
  gives  the  Dirac point but 
    the metallic state is  due to the over-tilted Dirac cone. 
Thus,  a  first-principles DFT calculation 
 was performed to show the Dirac electron  with 
  a small  gap $\sim 2$meV due to the spin-orbit coupling (SOC) 
and an effective model with transfer energies 
 was derived~\cite{Tsumuraya_Suzumura}.  
Because of the SOC-induced gap, spin Hall conductivity and anomalous spin 
conductivity under  magnetic field are 
 theoretically expected~\cite{Ogata_Matsuura}. 
 With such a TB model with complex transfer energies 
 and  the e-p interaction,
 it was shown that the conductivity at high temperatures is  almost constant  
 due to the increase  in  
the scattering by the phonon~\cite{Suzumura_Tsumuraya_2021}.  
 At low temperatures,  the   increase in resistivity was examined  
 by introducing the on-site Coulomb interaction~\cite{Kobayashi_2020},  
 which is treated with Hartree's approximation 
 with  the mean field  resulting 
 in  a second-order phase transition (at $T_c \sim$  0.003 eV) to a state with 
  a spin-ordered massive Dirac electron  with a gap. 
 It was pointed out that  such a gap competes with that induced by  
 SOI. 
It is of interest to verify experimentally
 such a magnetic ordered state~\cite{Fujiyama}.  
Further, 
 with an {\it ab initio} and two-dimensional
  extended Hubbard model treated within the 
 Hartree-Fock approximation~\cite{Kobayashi_2022},      
 the metal-insulator  crossover at $\sim$ 50 K,   followed by 
 the ordered states of a quantum spin  Hall insulator, was  shown, where 
 the SOI enhances  the resistivity  at low temperatures. 

In addition to the conductivity, there are several studies on 
  the Seebeck coefficient of  organic conductors. 
The formula for the Seebeck effects was established  using  linear 
 response theory~\cite{Kubo_1957,Luttinger1964,Fukuyama2024,Ogata_Fukuyama}.  
  For \ET, the coefficient was observed  experimentally under  hydrostatic  pressure,~\cite{Tajima_Seebeck,Konoike2013}, which is 
   measured along   the $x$ direction (perpendicular to the molecular 
  stacking axis)~\cite{Tajima_Seebeck}. 
  The calculation of the Seebeck coefficient $S_{\nu}$ was performed  along the $y$ direction~\cite{Kobayashi_2020_ET},  
and  along the $x$ direction~\cite{Suzumura_PRB_2023}, (i.e.,  the same  direction as that in an experiment).   
With decreasing temperature,  the former at ambient pressure  
 shows a  decrease in  $S_{y}$  by  accounting  for 
the correlation, while the latter under uniaxial pressure 
 shows a sign change and a minimum at low temperatures. 
Although the case of uniaxial pressure shows $S_x > 0$ at high temperatures  
  and  exhibits qualitatively the same  $T$ 
 dependence as that of the experiment, 
 a  problem still remains for the case of hydrostatic pressure. 
 A model calculation using the transfer energies  with the extended H\"uckel method~\cite{Kondo2009} with  a previous first-principles calculation at ambient pressure\cite{Kino2006}  gave  
 a negative  Seebeck coefficient ($S_x < 0$)~\cite{Suzumura_PRB_2023},   whereas experimental results showed   a positive Seebeck coefficient at finite temperatures.\cite{Tajima_Seebeck,Konoike2013}
 Then, with  a TB model obtained  from an $ab$ $initio$ calculation 
 with  the experimentally obtained crystal structure 
 at 1.76 GPa,~\cite{Kondo2009}
 it was shown that   the Seebeck coefficients 
 in \ET \; under hydrostatic pressure  are  positive 
 in both $x$ and $y$ directions~\cite{Suzumura_JPSJ_2024}
  indicating the hole-like behavior. 
 To  understand a competition between the conduction and  valence bands, 
  a  method 
 at low temperatures was proposed  in which 
 the sign of $S$  is determined by that of    the energy derivative of 
 the spectral conductivity $\sigma_\nu(\ep,T)$  
with respect to $\ep$ at a chemical potential~\cite{Suzumura_PRB_2023}.  
It was recently reported\cite{Furukawa_JPS} that the Seebeck coefficient
  in \BETS\ shows
 a noticeable  difference from that in \ET; 
 i.e., $S_x$ is negative in contrast to $S_x>0$  
in $\alpha$-(BEDT-TTF)$_2$I$_3$,  and 
$S_y$ changes sign from negative to positive as the temperature increases. 
It is important to clarify the origin of these differences between 
$\alpha$-(BETS)$_2$I$_3$ and $\alpha$-(BEDT-TTF)$_2$I$_3$.

In this paper, we  theoretically  study the Seebeck coefficient of \BETS \;  
   using  a TB model  with transfer energies, which are   
 derived from the  first-principles relativistic DFT
  calculation~\cite{Tsumuraya_Suzumura, Suzumura_Tsumuraya_2021}. 
The present paper is organized as follows.  In Sec. II, a formulation 
  for $\alpha$-(BETS)$_2$I$_3$ is given, in which  a TB model consists of 
 transfer energies with   both real and imaginary parts.  
     The electronic states  of the TB model are shown using  
 the conduction and valence energy bands around the Dirac point 
 and the  density of states. 
 In Sec. III, the Seebeck coefficient $S$ is calculated 
 by adding  the impurity and e-p scatterings.
 The   $T$ dependence  of  $S$ for both the $x$ and $y$ directions 
  is shown  and   analyzed in terms of  
   the  spectral conductivity with velocities. 
 Section IV is devoted to a summary and discussion.

\section{Model and Electronic States}
\subsection{Model Hamiltonian} 
\begin{figure}
  \centering
\includegraphics[width=6cm]{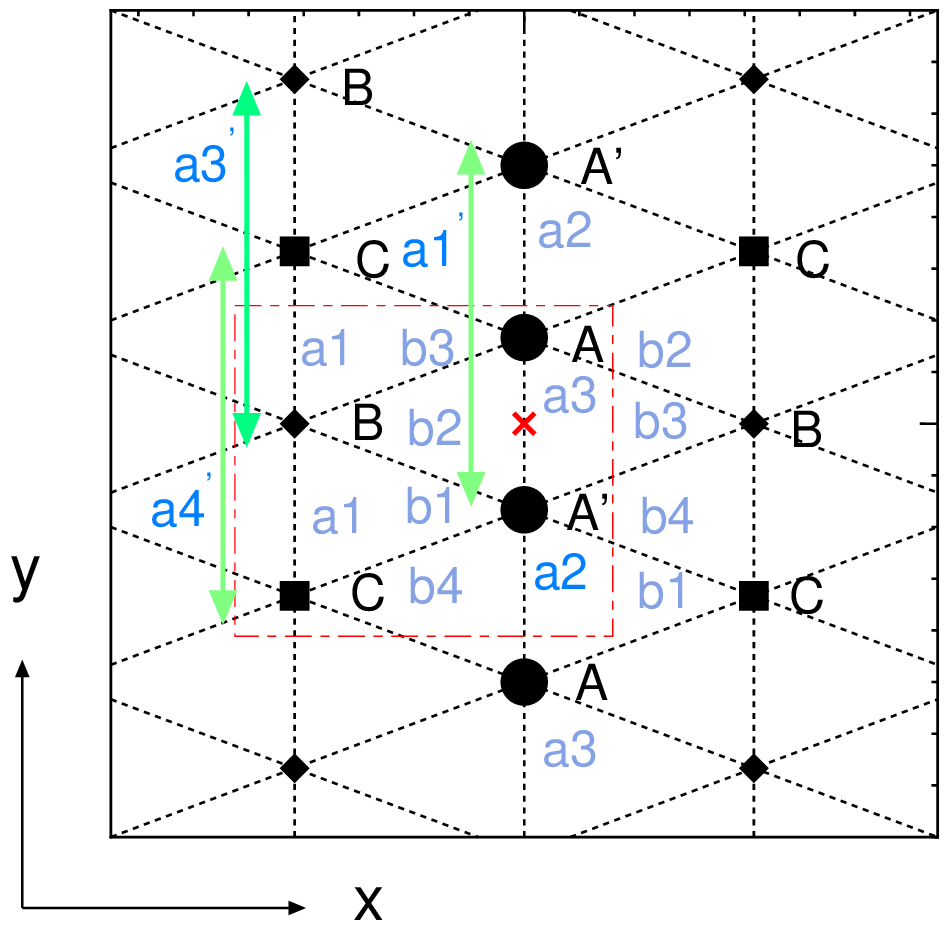}
     \caption{(Color online)
Crystal structure of \BETS.
There are four molecules, $A$, $A'$, $B$ and $C$ 
 in the unit cell (dot-dashed line), 
which forms a square lattice.
 The transfer energies with the same spin (opposite spin )
 are shown  in Table \ref{table_1} (Table \ref{table_2}); 
 the energies  for the nearest-neighbor (NN) sites 
 are given by $a1, \cdots, b4$.
Those  for the next-nearest-neighbor (NNN) sites 
 with the same molecules  are given by 
 $a1'$, $a3'$, and $a4'$   along the $k_y$ direction 
and $s1, \cdots, s4$ (not shown here) along the $k_x$ direction.  
Those  for NNN sites with different kinds of molecules 
 are given by $c1, \cdots, c4$ and  $d0, \cdots, d3$, which are not  shown 
 here, but we refer readers to our previous paper~\cite{Tsumuraya_Suzumura}.
The cross denotes an inversion center between $A$ and $A'$. 
$x$ ($y$) denotes a coordinate  perpendicular  
to (along) the molecular stacking  direction. 
 }
\label{fig1}
\end{figure}

We consider a two-dimensional Dirac electron system, 
 which is given by~\cite{Suzumura_Tsumuraya_2021} 
\begin{equation}
H = H_0  + H_{\rm p} +  H_{\rm e-p} +H_{\rm imp} \; . 
\label{eq:H}
\end{equation}
Here, $H_0$ describes a TB model  of 
 the  organic conductor, \BETS, consisting of four  molecules 
per unit cell (Fig.~\ref{fig1}). 
 The second term  denotes   the harmonic phonon  given by  
 $H_{\rm p}= \sum_{\bq} \omega_{\bq} b_{\bq}^{\dagger} b_{\bq}$ 
 with $\omega_{\bq} = v_s |\bq|$ and  $\hbar$ =1. 
 The third term is  the electron--phonon (e--p) interaction 
with a coupling constant $g_{\bq}$,
\begin{equation}
 H_{\rm e-p} = \sum_{s= \pm}\sum_{\bk, \g} \sum_{\bq}
   g_{\bq} c_{\g s}(\bk + \bq)^\dagger c_{\g s}(\bk) 
(b_{\bq} + b_{-\bq}^{\dagger})
\; .
\label{eq:H_e--p}
\end{equation}
 $c_{\g s}(\bk) = \sum_{\alpha} d_{\alpha \g}a_{\alpha s}(\bk)$ 
 with spin  $s$, which  is obtained by diagonalizing  $H_0$ as shown later. 
 The e--p scattering is considered 
 within  the same band (i.e., intraband). 
The last term of Eq.~(\ref{eq:H}), $H_{\rm imp}$, denotes a normal  impurity 
 scattering.

The TB model, $H_0$, is expressed as 
\begin{eqnarray}
H_0 &=& \sum_{i,j = 1}^N \sum_{\alpha, \beta = 1}^4 \sum_{s, s'= \pm}
 t_{ij; \alpha s,\beta s'} a^{\dagger}_{i \alpha s} a_{j \beta s'} 
\; , 
\label{eq:Hij}
\end{eqnarray}
where   $a^{\dagger}_{i \alpha s}$ denotes a creation operator 
 of an electron 
 at  the  molecule $\alpha$ 
 [= A(1), A'(2), B(3), and C(4)] in the unit cell 
  with  the $i$-th lattice site.
$s=+$ and $s=-$ denote $\uparrow$ and $\downarrow$  spins, respectively. 
 $N$ is the total number of square lattice sites and 
 $t_{ij; \alpha s,\beta s'}$ are the transfer energies 
for  the nearest  neighbor (NN) and next-nearest 
 neighbor (NNN) sites.~\cite{Tsumuraya_Suzumura} 
$H_0$ is rewritten using 
  a Fourier transform of the operator $a_{j,\alpha,s}$, which  
 is given by   
 $a_{j,\alpha, s} = 1/N^{1/2} \sum_{\bk} a_{\alpha s}(\bk) \exp[ i \bk \cdot \bm{r}_j]$.
 The wave vector  $\bk = (k_x,k_y)$ 
 is taken within  $\bG$, which  
 denotes  a reciprocal lattice vector of the square lattice. 
 The quantity $\bG$/2 
  corresponds  to the vector of the time-reversal invariant momentum (TRIM).
 Thus, $H_0$ is rewritten  as 
\begin{subequations}
\begin{eqnarray}
  H_0 = \sum_{\bk} \sum_{s, s'} \sum_{\alpha, \beta}  
      h_{\alpha \beta; s s'}(\bk)  a^{\dagger}_{\alpha s}(\bk) 
         a_{\beta s'}(\bk)\; . 
\label{eq:H_k}
\end{eqnarray}
 The transfer energy $h_{\alpha \beta; s s'}(\bk)$
 is expressed as 
\begin{eqnarray}
   h_{\alpha \beta; s s'}(\bk) = 
   \sum_{ j (\not= i)} t_{ij; \alpha s,\beta s'}
 \exp [-i \bk \cdot({\br}_i-{\br}_j)]
   \; ,
\label{eq:def_transfer}
\end{eqnarray}
\end{subequations}
where site potentials are added as shown in  Appendix A. 
In Eqs.~(\ref{eq:H_k})and (\ref{eq:def_transfer}), 
 the SOC is included  due to the $d$  electrons in  Se atom.

\subsection{Transfer energies of TB model}
The transfer energy $t_{ij; \alpha s,\beta s'}$ in Eq.(\ref{eq:Hij}) is derived from first-principles DFT calculations.~\cite{Tsumuraya_Suzumura} It is given by
\begin{eqnarray}
t_{\alpha \beta; s s'}(\mathbf{R})
=\langle\phi_{\alpha s,0}|H_{k}|\phi_{\beta s',\mathbf{R}} \rangle ,
\label{equ_t}
\end{eqnarray}
where $H_{k}$ is the one-body part of the \textit{ab initio} Hamiltonian for \BETS~\cite{Kino2006}.  $\phi_{\alpha s,\mathbf{R}}$ denotes the maximally-localized Wannier function (MLWF) spread over the BETS molecule, with $\alpha s$ representing distinct MLWFs. 
$\mathbf{R}$ indicates the location of the $j$-th unit cell relative to the $i$-th unit cell. 
The crystal structure is based on the experimental structure at 30 K and ambient pressure~\cite{Kitou2021}, with hydrogen positions optimized using first-principles methods.

As shown in Eq.(\ref{equ_t}), $t_{ij; \alpha s,\beta s'}$ depends solely on the relative positions of the $i$-th and $j$-th sites. Using Bloch functions obtained from DFT calculations\cite{HK1964, KS1965}, we constructed the MLWFs with the \texttt{wannier90} code~\cite{Marzari1997, Isouza2001}.
Each MLWF is centered at the midpoint of the central C=C bonds in the BETS molecule.
To construct MLWFs, we selected eight bands near the Fermi level (four bands consisting of the highest occupied molecular orbital of 
 BETS molecule with up and down spins).
The DFT band structure including SOC was calculated using fully relativistic pseudopotentials with plane-wave basis sets in 
 QUANTUM  ESPRESSO~\cite{PAW1994,Giannozzi,QE}.  
For the exchange-correlation functional, we used the Perdew-Burke-Ernzerhof 
 generalized gradient approximation~\cite{GGAPBE}. 
 The fully relativistic pseudopotentials were
generated using atomic code (version 6.3)~\cite{Atoms} with a
pseudization algorithm proposed by Troullier and Martins~\cite{TM}.
The valence configurations were (1$s$)$^{1}$ for H,
 (2$s$)$^2$(2$p$)$^2$ for C, 
(3$s$)$^2$(3$p$)$^4$ for S, (4$s$)$^2$(4$p$)$^4$(3$d$)$^{10}$ for Se, 
and (5$s$)$^2$(5$p$)$^5$(4$d$)$^{10}$ for I atoms.
The cutoff energies for plane waves and charge densities were 48 and 488 Ry, respectively. 
For the self-consistent loops, 
 we employed a 4 $\times$ 4 $\times$ 2 uniform $\bk$-point mesh 
 with Gaussian smearing. 
The computational details were provided in our previous work~\cite{Tsumuraya_Suzumura,Suzumura_Tsumuraya_2021}. 

In Tables \ref{table_1} and  \ref{table_2}, the transfer energies 
 $w_h$ obtained from Eq.~(\ref{equ_t}) are shown, 
 where   $w_h (s = s')$ and  $w_h (s = -s')$ correspond  to  
the same spin  and opposite spin,  respectively.
 In Table \ref{table_1}, the NN 
  interactions are labeled  $a_1, \cdots, b_4$, 
 while the NNN interactions 
 are labeled  
 $a1', \cdots, a4'$, 
 $c1, \cdots, c4$, 
 $d0, \cdots, d3$,  
  and $s1, \cdots, s4$~\cite{Tsumuraya_Suzumura,Suzumura_Tsumuraya_2021}. 
The quantities $V_{\alpha}$ represent  the on-site potential energies. Since the  origin of the energy is arbitrary, we measure the on-site potentials $V_{\rm B}$ and $V_{\rm C}$ relative to those of $A$ and $A'$ (Appendix A). 
All energies are expressed in eV.

\begin{table}[b]
\centering
\caption{
Effective transfer energies $w_h$ $(s =s')$ of the same spin 
 and site-potential energies for \BETS. 
$\Delta V_{B}$ and $\Delta V_{C}$ are  site-potential energies of $B$ and $C$ molecular sites relative to $A$ (and $A^{\prime}$) site, respectively.  The definitions  are shown  in Eq.~(\ref{equ_t}) and Appendix A. 
 $t_{\alpha s,\beta,s'} \rightarrow w_h(s= s')$ 
 includes SOC and are labeled   
 $h = a1, \cdots, s4$. 
Note that Im[$w_h(s=\pm s')$] is replaced by  sgn(s)Im[$w_h(s=\pm s')$]. 
}
\label{table_1}
\begin{center}
\begin{tabular}{crr}
\hline\hline
$w_h$ $(s =s')$ &  Re($w_h$) &  Im($w_h)$ \\ 
\hline
$a1$  & 0.0053  & 0.00130  \\
$a2$  & -0.0201  &  0 \\
$a3$  & 0.0463  & 0 \\
\hline        
$b1$  & 0.1389  & 0.00674 \\
$b2$  & 0.1583  & 0.00731 \\
$b3$  & 0.0649  & 0.00203 \\
$b4$  & 0.0190  & -0.00120 \\
\hline        
$a1'$ & 0.0135  & 0 \\
$a3'$ & 0.0042  & 0 \\
$a4'$ & 0.0217  & 0 \\
\hline        
$c1$  & -0.0024  & -0.00056 \\
$c2$  &  0.0063  & -0.00010 \\
$c3$  & -0.0036  & -0.00040 \\
$c4$  & 0.0013   &  0.00027 \\
\hline        
$d0$  & -0.0009  & 0 \\
$d1$  & 0.0104   & 0 \\
$d2$  & 0.0042   & -0.00009 \\
$d3$  & 0.0059   & -0.00003 \\
\hline        
$s1$  & -0.0016  & -0.00017 \\
$s3$  & -0.0014  &  0 \\
$s4$  &  0.0023  &  0 \\
\hline
\hline
$\Delta V_B$ & -0.0047 & \\ 
$\Delta V_C$ & -0.0092 & \\   
\hline\hline
\end{tabular}
\end{center}
\end{table}

\begin{table}[b]
\centering
\caption{
Effective transfer energies  $w_h$ $(s =-s')$ of opposite spin 
 for \BETS. 
Quantities $w_h(s= -s')$, which    
 include  SOC and are labeled  
$h = b1_{so1}, \cdots, c4_{so2}$,~\cite{Tsumuraya_Suzumura}, 
 are estimated by averaging 
two components of spin freedom.
}
\centering
\label{table_2}
\begin{tabular}{crr}
\hline
\hline
$w_h$ $(s =-s')$ & Re($w_h$) & Im($w_h$) \\
\hline
$b1_{so1}$  & -0.0020 & 0.00077 \\
$b1_{so2}$  &  0.0020 & -0.00077 \\
$b2_{so1}$  & -0.0019 & -0.00017 \\
$b2_{so2}$  &  0.0019 &  0.00017 \\
$b4_{so1}$  &  0.0008 & -0.00097 \\
$b4_{so2}$  & -0.0008 &  0.00097 \\
$c1_{so1}$  &  0.0007 & -0.00033 \\
$c1_{so2}$  & -0.0007 &  0.00033 \\
$c2_{so1}$  &  0.0003 &  0.00022 \\
$c2_{so2}$  & -0.0003 & -0.00022 \\
$c3_{so1}$  &  0.0006 &  0.00018 \\
$c3_{so2}$  & -0.0006 & -0.00018 \\
$c4_{so1}$  &  0.0001 & -0.00008 \\
$c4_{so2}$  & -0.0001 &  0.00008 \\
\hline
\hline
\end{tabular}
\end{table}

\subsection{Electronic states} 
There are three kinds of SOC, which  act  on the same spins ($s = s'$) and
    the opposite spins ($s = -s'$) with two components of spin freedom. 
 The former case (the latter case) is denoted SOC I (SOC II) and is 
 treated 
 by a spin-decoupled  $4 \times  4$ matrix Hamiltonian 
 (an   $8 \times 8 $ matrix Hamiltonian
  coupled by up and down  spins).
  Further,  SOC III is defined  
 as  the case with only the real part of  SOC I, 
 which  also includes SOC. 

 The band energy of a TB model $H_0(\bk)$ 
 is obtained as   
\begin{eqnarray}
\label{eq:eigen_eq}
\sum_{\beta, s'} h_{\alpha\beta; s s'}(\bk) d_{\beta s' \g}(\bk)
   &=& E_{\g}(\bk) d_{\alpha s \g} (\bk)  \; , 
 \label{eq:E_alpha}
\end{eqnarray}
where $h_{\alpha\beta; s s'}(\bk)$ with the same spin 
 is shown   in  Appendix A. 
 $E_\gamma(\bk)$, with $\gamma = 1, \cdots, 4$ 
  for the spin-decoupled  Hamiltonian 
 and $\gamma = 1, \cdots, 8$ for  the spin-coupled Hamiltonian,  
  denotes the band energy in descending order.
 The Dirac point $\bkD$ is obtained  
from a minimum of $E_c(\bk) -  E_v(\bk) ( >  0)$, where 
  $E_c(\bk)$ and $E_v(\bk)$ are the  conduction   
 and  valence bands above and below a chemical potential, respectively. 

  From the three-quarter-filled condition, the chemical potential 
 $\mu$ is calculated by 
\begin{eqnarray}
  \frac{1}{N(2N)} \sum_{\bk} \sum_{\gamma = 1}^{4(8)}  f[E_{\gamma}(\bk)-\mu]=  3 \; ,  
  \label{eq:mu}
\end{eqnarray}
where  
 $f(\ep)= 1/[\exp[\ep/k_{\rm B}T]+1]$ with $T$ being temperature 
 and the Boltzmann constant taken as $k_{\rm B }= 1$. 
 Thus,   $E_1(\bk)$ and $E_2(\bk)$   for the $4 \times 4$ matrix
 and $E_2(\bk)$ and $E_3(\bk)$ for the $8 \times 8$ matrix   
 give the conduction and valence bands, 
 respectively.  
 The DOS per spin 
  is calculated from 
\begin{eqnarray}
  D(\omega) = \frac{1}{N(2N)} \sum_{\bk} \sum_{\gamma =1}^{4(8)}  
 \delta [\omega - E_{\gamma}(\bk)].
  \label{eq:DOS}
\end{eqnarray}

\begin{figure}
  \centering
\includegraphics[width=6cm]{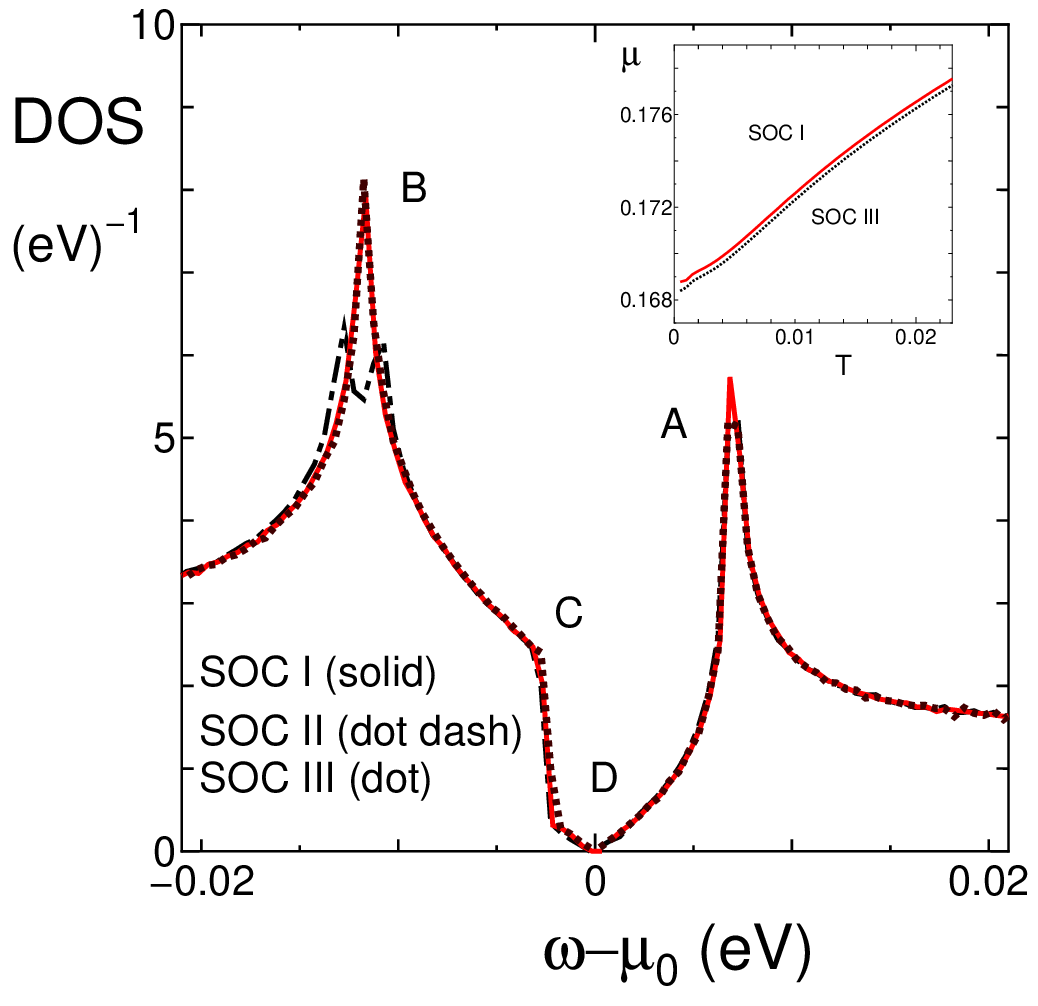}
     \caption{(Color online )
 DOS per spin as a function of $\omega - \muz$, 
  where $\muz=0.1687$ denotes  $\mu$ at $T = 0$. 
The solid line (SOC I) shows DOSs with the 
 same spin ($s = s'$), 
  while the dotted line (SOC III) 
   denotes  DOSs with  only  Re[$w_h(s = s')$].
 The  dash-dotted  line (SOC II) shows DOSs in the presence of  
 SOC with both the same and opposite 
spins ($s = s'$ and $s = -s'$). 
  The energy at the Dirac point is shown by $D$ and those of  
 the Van Hove singularity are shown by  $A$ and $B$, which are 
 explained in the text.
 The inset denotes $\mu$ as a function of  $T$. 
}
 \label{fig2}
\end{figure}

In Fig.~\ref{fig2}, DOS is shown  as a function of $\omega - \muz$
with  $\muz$ being $\mu$ at $T = 0$.
 The solid line (SOC I)  shows the DOS for the same spin ($s = s'$)
 with both Re[$w_h(s = s')$] and Im[$w_h(s = s')$],   
  while the dotted line (SOC III)  
   denotes  DOS only for Re[$w_h(s = s')$].
The gap on the Dirac point 
 is found to be   
$2.1 \times 10^{-3}$ for SOC I (solid line) 
  but zero for  SOC II (dot-dashed line) and SOC III (dotted line).
 Thus,  the former transfer energy gives rise to  the  effect of SOC 
   on the transport. 
 The  dot-dashed  line (SOC II) shows the DOS in the presence of  
 both the same and opposite 
   spins ($s = s'$ and $s = -s'$). 
  For $\omega - \muz \simeq -0.01$, 
 the DOS of SOC I and SOC III show a single peak,  
 but the DOS of SOC II shows 
 a double peak due to 
 the splitting   
 by the coupling between the  opposite spins.
There is an asymmetry of the DOS around $\omega = \muz$.  
 The DOS for the conduction band ($\omega > \muz$)  is  
  smaller than that for the valence band ($\omega < \muz$) 
  because of the presence of the shoulder $C$. 
Thus, $\mu$ at finite temperatures becomes larger  than $\muz$.  
The inset in Fig.~\ref{fig2} shows the $T$ dependence of the chemical potential $\mu$, where
 $\mu$ for SOC I is slightly larger than that for SOC III. 
With increasing $T$, $\mu$ 
increases monotonously, and we find that  the relation
   $\mu(T) - \muz \sim 0.4 T$ approximately  holds.
 
\begin{figure}
  \centering
\includegraphics[width=8cm]{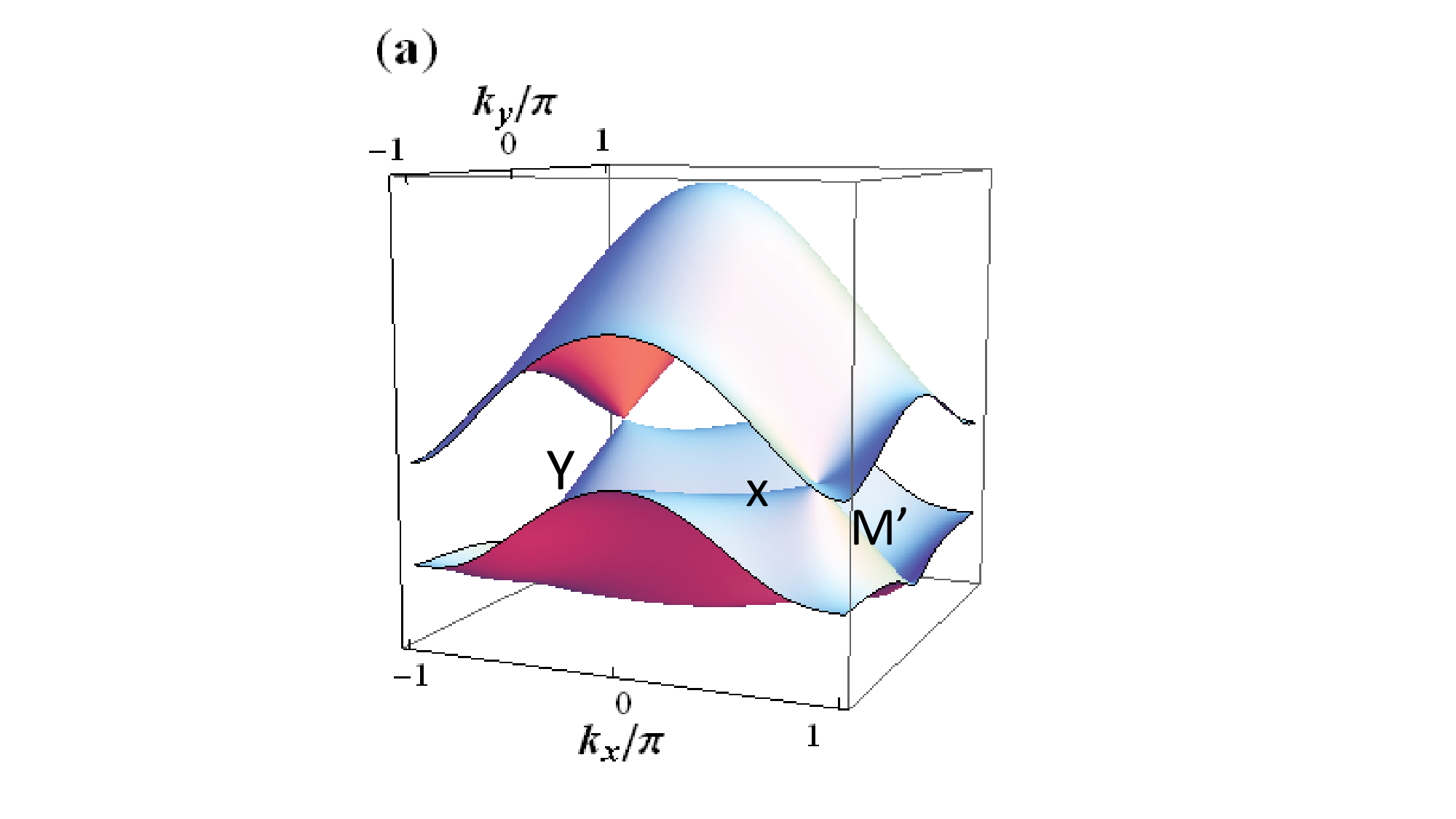}\\
\includegraphics[width=3.5cm]{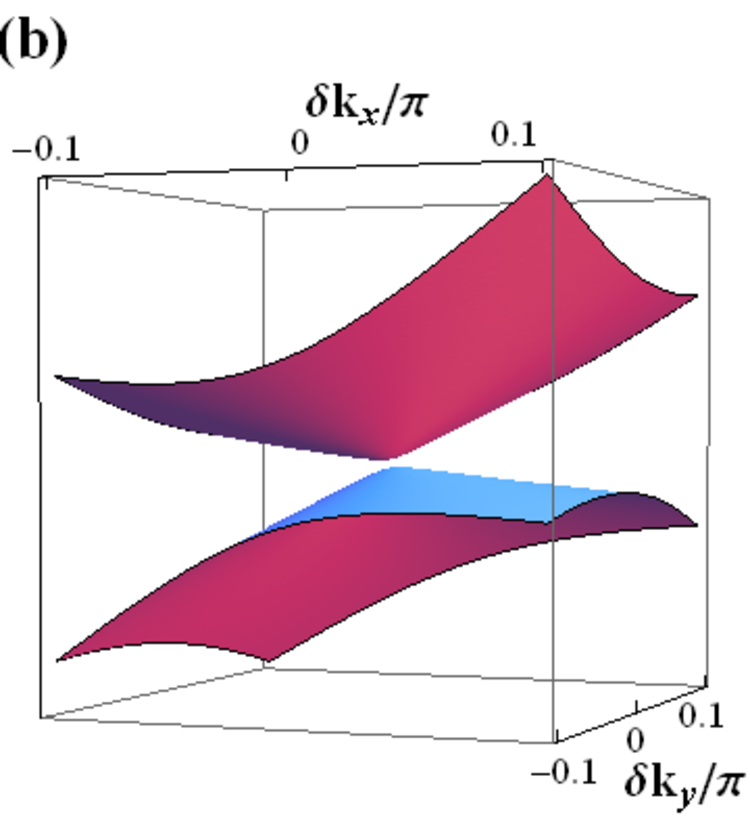}
\includegraphics[width=4.5cm]{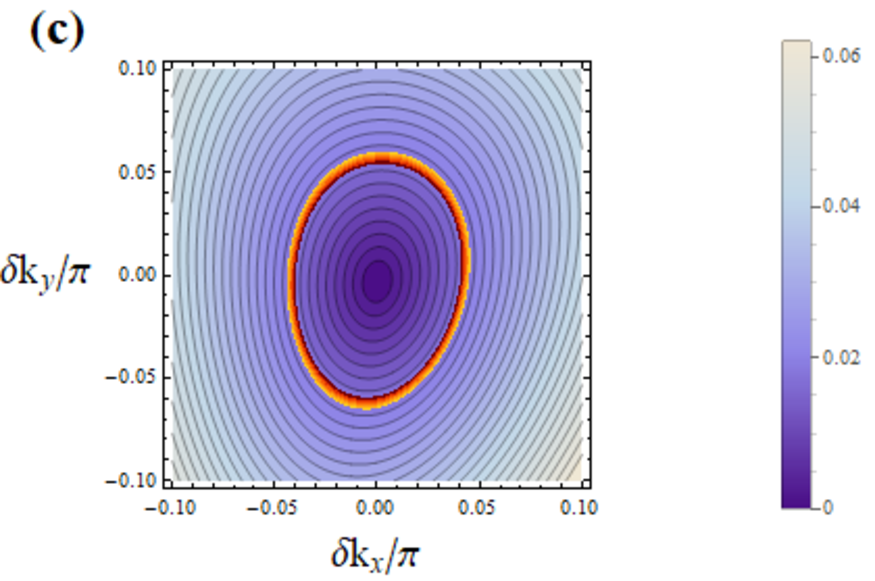}\\
\includegraphics[width=6.5cm]{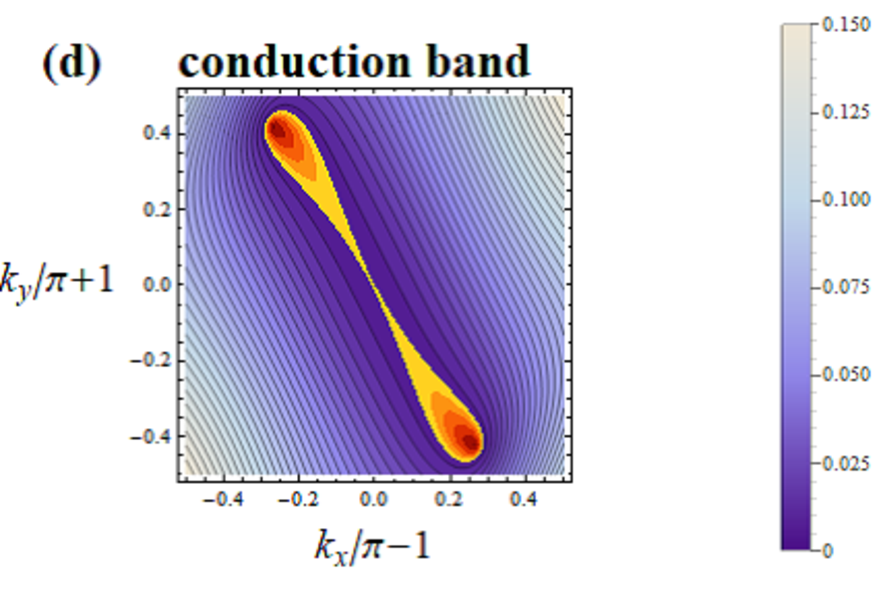}\\
\includegraphics[width=6.5cm]{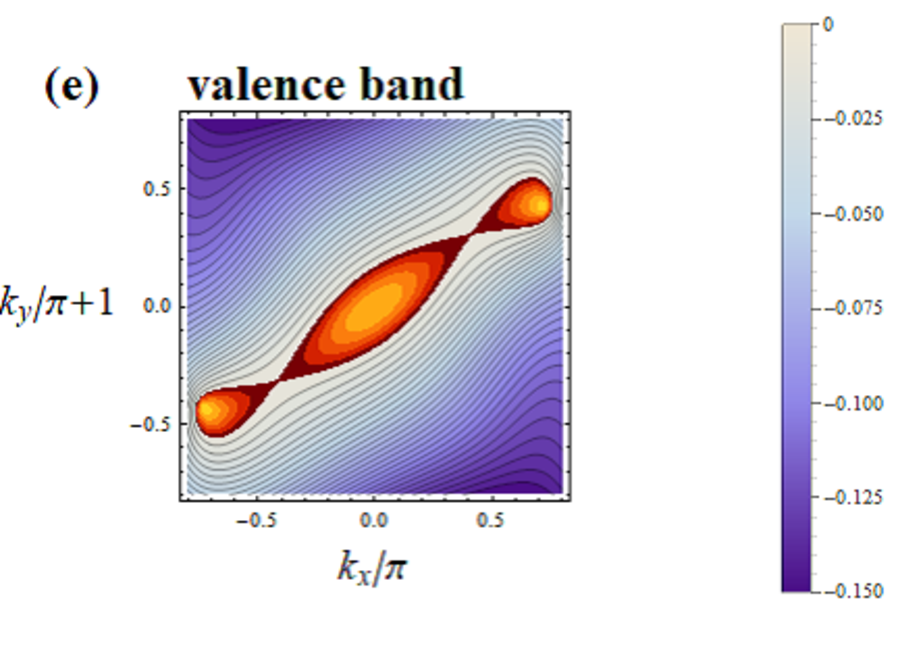}\\
     \caption{(Color online )
(a) Two bands, conduction and valence bands, given by  
  $E_1(\bk)$ (upper band) and $E_2(\bk)$ (lower band), 
 which contact at a Dirac point.
 TRIMs Y and M' denote  $(k_x,k_y)/\pi$ = (0,-1) and (1,-1) respectively.
 The cross in $E_2(\bk)$ shows a saddle point on a line connecting Y and the Dirac point.
 (b) Magnified two bands 
      $ E_1(\bk)$ and $E_2(\bk)$  
       showing a tilted Dirac cone around the Dirac point $\bkD$,  
 where   $\delta \bk = \bk - \bkD$ 
   and  $\bkD = -(0.729,-0.577)\pi$.  
 (c) Contour plots of   $E_1(\bk) - E_2(\bk)$ on the plane of  $\delta \bk$, 
     where  the red line corresponds to 
    $E_1(\bk) - E_2(\bk) =  0.02$.
(d) Contour plots of conduction band $E_1(\bk)$ 
 near the M' point $(k_x, k_y) = (\pi, -\pi)$. 
 The yellow to red contours show the region with $0 \le E_1 - \muz < 0.0074$. 
 The Dirac points are located at $(k_x/\pi -1, k_y/\pi+1) = \pm(-0.271,0.423)$, and 
 the pinch point indicates the saddle point at M'.
 (e)  Contour plots of valence band $E_2(\bk)$
 near the Y point $(k_x, k_y) = (0, -\pi)$. 
 The yellow to red contours show the region with $-0.014 \le E_2 - \muz \le 0$. 
 The Dirac points are located at $(k_x/\pi, k_y/\pi+1) = \pm(0.729,0.423)$.
 The energy at Y point has a maximum and there are two saddle points, which are
 represented by two pinch points.
}
 \label{fig3}
\end{figure}

  Hereafter, we examine the case of SOC  with the same spin 
  (intra-SOC), i.e., 
 the $4 \times 4$ matrix Hamiltonian.  
  Using   the TB model with transfer energies given by Table \ref{table_1},
 we examine energy bands  obtained from Eq.~(\ref{eq:E_alpha}). 
Figure \ref{fig3}(a) shows  
 conduction and valence bands given by  
$E_1(\bk)$ (upper band) and $E_2(\bk)$ (lower band).
 There is a Dirac point $\bkD$ where  a gap 
 corresponding to a minimum of 
 $E_2(\bk)- E_1(\bk)$ is found at  $\bk = \bkD$.
 Two Dirac points  exist at $\bkD$ and $- \bkD$, 
 with $\bkD = -(0.729,-0.577)\pi$ 
 in the second and fourth quadrants, respectively  on the the $k_x-k_y$ plane. 
 (We set the length of the unit cell equal to 1.)
 Figure \ref{fig3}(b) shows a  magnified scale of 
 the  conduction and valence bands  $ E_1(\bk)$ and $E_2(\bk)$ 
  around $\bkD$  
 as a function of  $\delta  \bk = \bk - \bkD$.  
  They are   described by 
 the Dirac cone, which  is tilted almost along the $k_x$ axis. 
   Note that the gap  ($\simeq$ 0.002) at $\bkD$ 
    comes from the SOC, which exists 
     in  
 transfer energies (Table \ref{table_1}).
 Figure \ref{fig3}(c) shows contour plots of 
  $E_1(\bk) - E_2(\bk)$, which is  
 the difference between the conduction and valence bands.
 The contour is similar but deviates from  the ellipsoid  with  
  the minor axis being close to the  $k_x$ axis. 
 The ellipsoid suggests  anisotropy  of the Dirac cone, leading to 
  the relation,    $v_x > v_y$ for the velocity and   
     $\sigma_x > \sigma_y$ for the electric conductivity. 
  Deviation from  the ellipsoid  results in 
 the Seebeck coefficient. 

Figure \ref{fig3}(d) shows
 contour plots of the  conduction band $E_1(\bk)$ 
 near the $M'$ point $(k_x, k_y) = (\pi, -\pi)$. 
 The Dirac points are located at $(k_x/\pi -1, k_y/\pi+1) = \pm(-0.271,0.423)$.
 The pinch point indicates the saddle point at $M'$, which gives  
 a Van Hove singularity at $A$ in Fig.~\ref{fig2}. 
 Figure \ref{fig3}(e) shows
 contour plots of the valence band $E_2(\bk)$
 near the $Y$ point $(k_x, k_y) = (0, -\pi)$. 
 The Dirac points are located at $(k_x/\pi, k_y/\pi+1) = \pm(0.729,0.423)$.
 There is a local maximum  at the $Y$ point corresponding to 
 the shoulder of $C$  in Fig.~\ref{fig2}. 
 There are two saddle points at $(k_x/\pi, k_y/\pi+1) \simeq \pm(0.40,0.40)$, 
  which are represented by two pinch points.
 They give  a Van Hove singularity  at $B$  in Fig.~\ref{fig2}.

\section{Seebeck Effects}

Using the  linear response theory,
~\cite{Kubo_1957,Luttinger1964}
 the electric current density 
$\bm{j} =(j_x,j_y)$ is obtained 
 from  the electric field $\bm{E} = (E_x, E_y)$
 and the temperature gradient $\nabla T$ as
\begin{eqnarray}
 j_{\nu} = L_{11}^{\nu} E_\nu  - L_{12}^{\nu} \nabla_\nu T/T
                                         \; ,  \quad {\rm }\ \nu=x, y\; .
   \label{eq:j}
\end{eqnarray}
The coefficients $L_{11}^{\nu}$ and $L_{12}^{\nu}$ denote 
the electric conductivity and 
the thermoelectric conductivity,  
 which are respectively written as 
~\cite{Ogata_Fukuyama,Fukuyama2024}
\begin{eqnarray}
L_{11}^{\nu} &=&  \sigma_{\nu}(T) = 
              \int_{- \infty}^{\infty} d \ep  
            \left( - \frac{\partial f(\ep - \mu) }{\partial \ep} \right)
\sigma_{\nu}(\epsilon,T)  \; ,
        \nonumber \\
    \label{eq:L11}
                                                \\
 L_{12}^{\nu} &=&  \frac{-1}{e}
              \int_{- \infty}^{\infty} d \ep  
  \left( - \frac{\partial f(\ep -\mu) }{\partial \ep} \right)
 (\ep - \mu) \sigma_{\nu}(\ep,T)   
  \; . \nonumber  \\
          \label{eq:L12}
\end{eqnarray}
 The electric conductivity per spin 
 is given by  $\sigma_\nu(T)$. 
 In Eqs.~(\ref{eq:L11}) and (\ref{eq:L12}), 
 $\sigma_{\nu}(\epsilon,T)$ 
 denotes the spectral conductivity, which is expressed 
 in terms of the conduction band ($\g =1$) and valence band ($\g=2$) 
~\cite{Suzumura_PRB_2023}: 
\begin{eqnarray}
\sigma_{\nu}(\epsilon,T) &=&   
      \sum_{\gamma, \gamma'=1}^2\sigma_{\nu}^{\gamma \gamma'}(\epsilon,T)
                     \; ,          
  \label{eq:spectral}
\end{eqnarray}
 where the component 
 is expressed as
\begin{eqnarray}
 \sigma_{\nu}^{\gamma \gamma'}(\epsilon,T)  & = &
 \frac{e^2 }{\pi \hbar N} 
  \sum_{\bk}
  v^\nu_{\gamma \gamma'}(\bk)^* 
  v^{\nu}_{\gamma' \gamma}(\bk)   \nonumber \\
      &  \times  &
     \frac{\Gamma_\g}{[\ep - E_{\gamma}(\bk)]^2 + \Gamma_\g^2} 
 \frac{\Gamma_{\g'}}{[\ep - E_{\gamma'}(\bk)]^2 +  \Gamma_{\g'}^2}
  \; ,  \nonumber \\
  \label{eq:spectral_comp}
\end{eqnarray}
 with $v^{\nu}_{\gamma \gamma'}(\bk)  =  \sum_{\alpha \beta}$
 $d_{\alpha \gamma}(\bk)^* 
   (\partial h_{\alpha \beta}/\partial k_{\nu})
 d_{\beta \gamma'}(\bk).$~\cite{Katayama2006_cond}  
 $h = 2 \pi \hbar$ and $e (>0)$ are  Planck's constant 
 and the electric charge,  respectively.  
It should be noticed that the quantity $\Gamma_\gamma$ in Eq.~(\ref{eq:spectral_comp}) comes from the damping of the electron by  both the impurity scattering and the inelastic electron-phonon scattering, as described below. The former contribution is separated from the latter one  within the lowest order of perturbation.
In the present numerical calculation,   $\sigma_\nu(T)$  
 and  $\sigma_\nu(\ep,T)$ are  normalized 
 by $\sigma_0 = e^2/(\pi h)$.

With $\sigma_{\nu}^{\gamma \gamma'}(\epsilon,T)$, 
 the Seebeck coefficient $S_\nu$ 
 is obtained as 
\begin{eqnarray}
S_{\nu}(T) = \frac{L_{12}^\nu}{T L_{11}^{\nu}}
    =\sum_{\g, \g'} S_{\nu}^{\g\g'}(T) \; ,
     \label{eq:S}
\end{eqnarray}
where the component is expressed as 
\begin{eqnarray}
 S_{\nu}^{\g\g'}(T) &=& \frac{1}{T L_{11}^{\nu}} 
                                     \nonumber \\
   \times  \frac{-1}{e} &&
              \int_{- \infty}^{\infty} d \ep  
  \left( - \frac{\partial f(\ep - \mu) }{\partial \ep} \right)
         (\ep - \mu)  \sigma_{\nu}^{\g\g'}(\ep,T)
                    \; .       \nonumber \\
   \label{eq:S_comp}
\end{eqnarray}
  Note that the diagonal components of 
 $(\g,\g')$=(1,1) and (2,2) give  the intraband contribution 
 and the off-diagonal components give 
 the interband contribution.  

To calculate the spectral conductivity in Eq.~(\ref{eq:spectral_comp}), 
 we need 
 the relaxation rate $\Gamma_\g$ of electrons 
 in the $\g$ band.
As in our previous paper\cite{Suzumura_PRB_2018}, we assume 
\begin{eqnarray}
\Gamma_{\g}  = \Gamma + \Gamma_{\rm ph}^{\g} \; ,
\end{eqnarray}
where $\Gamma$ comes from the impurity scattering 
and $\Gamma_{\rm ph}^\g$ comes from  the phonon scattering
 \suzu{obtained from the lowest-order 
 self-energy of electrons with the electron-phonon coupling 
 $g_{\bq}$:
}
\begin{subequations}
\begin{eqnarray}
  \Gamma_{\rm ph}^\g &=& C_0R  T|\xi_{\g,\bk}|
  \; ,
 \label{eq:eq16a}
        \\ 
R &=& \frac{\lambda}{ \lambda_0}
 \; ,  
 \label{eq:eq16b} 
\end{eqnarray}
 \end{subequations}
 where $\lambda = |g_{\bq}|^2/\omega_{\bq}$, 
    $\xi_{\g,\bk} = E_{\g}(\bk) - \mu$, 
    $C_0 = 6.25\lambda_0/(2\pi v^2)$ 
   and $\lambda_0/2\pi v = 0.1$.  
  $v \simeq 0.05$~\cite{Katayama_EPJ} 
 denotes the averaged velocity of the Dirac cone. 
 $\lambda_0$ corresponds to  $\lambda$
  for an organic conductor\cite{Rice,Gutfreund}, 
and  
  $\lambda$ becomes  independent of $|\bq|$  for small $|\bq|$. 
  $R$ is taken as a parameter. 
   We take mainly $\Gamma = 0.0005$ and  $R =1$  in the present numerical 
 calculation~\cite{Suzumura_JPSJ2021,Suzumura_PRB_2018}.   
 
\begin{figure}
  \centering
\includegraphics[width=6cm]{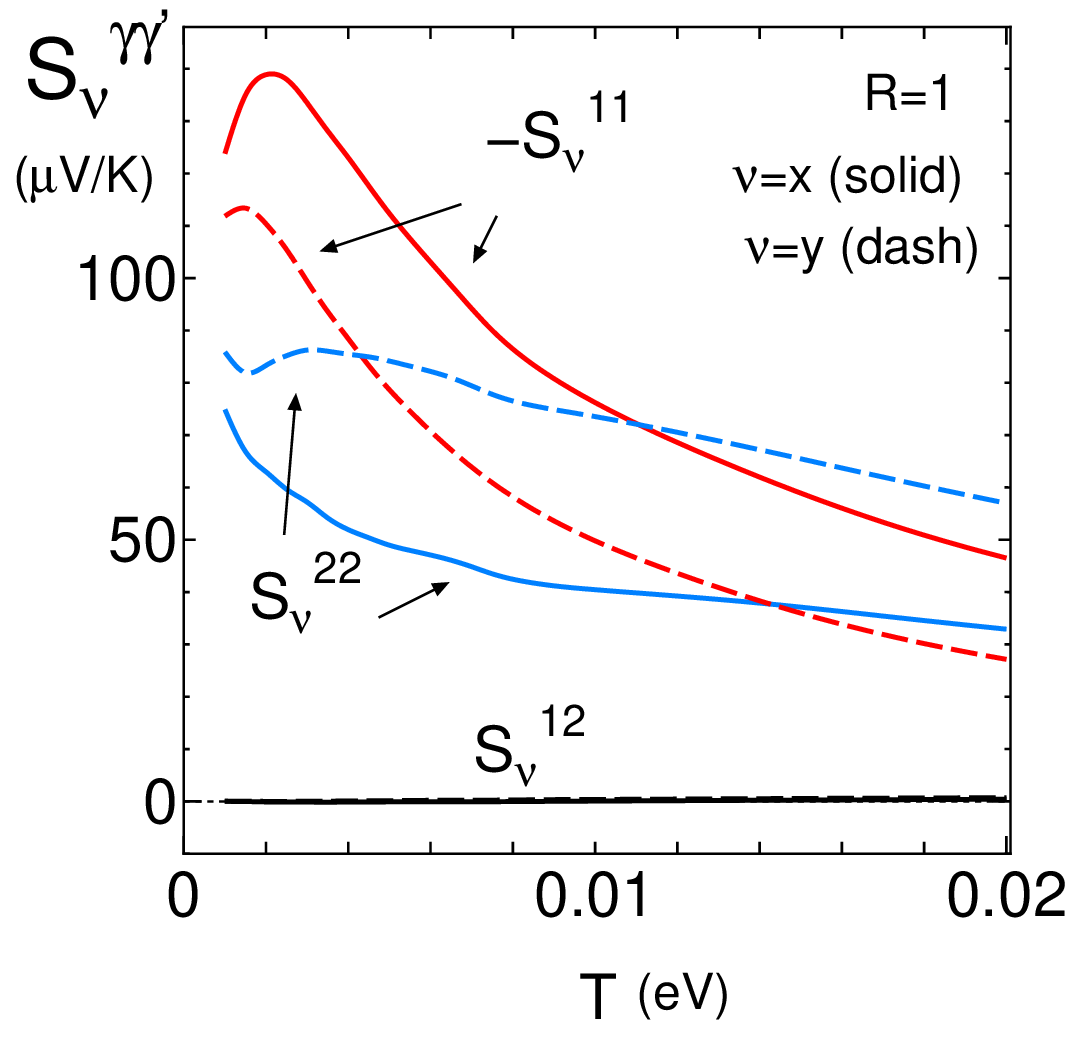}
     \caption{(Color online )
$T$  dependence of  the components 
 of the Seebeck coefficient $S_\nu^{\g \g'}$ 
 given by Eq.~(\ref{eq:S_comp}).
The  normalized e--p  interaction is taken as $R$ = 1,
 which corresponds to  weak coupling. 
 The solid  and dashed lines  correspond  
 to $\nu = x$ and $y$, respectively.
}
\label{fig4}
\end{figure}

  First,  the sign of  the Seebeck coefficient is examined by 
 noting that both 
 the conduction and valence bands contribute to  
 the Seebeck coefficient  as seen from Eqs.~(\ref{eq:spectral_comp}), 
(\ref{eq:S}) and (\ref{eq:S_comp}). 
 Since  the intraband  contribution of Eq.~(\ref{eq:S_comp}) gives 
 $S_{\nu}^{\g \g} < 0 (> 0)$ for $\ep - \mu > 0 (<0)$,  
  the  conduction (valence) band gives the negative (positive) sign  
  for the component of the Seebeck coefficient.
Therefore, the sign of the Seebeck coefficient is determined 
 by the competition  between the conduction and valence bands.  
 Figure \ref{fig4}  shows  the $T$ dependence of 
 $S_{\nu}^{\g\g'}$, 
  where the component of $(\g,\g')$ = (1, 1) is negative due to the 
 conduction band (i.e.,  electrons) and  
 that of (2,2) is positive due to  the valence band (i.e., holes). 
 The case of  $\nu = y$  shows   
 $S_{y}^{22} < - S_{y}^{11}$  for $T <  0.004$ 
 and 
  $S_{y}^{22} > - S_{y}^{11}$ for  $T >  0.004$. 
 For the $y$ direction, a crossover of the dominant contribution 
occurs  from the electron to the hole   with increasing $T$, while 
 the dominant contribution in  the case of  $\nu = x$ is always 
 the electron, i.e.,     
  $S_{x}^{22}  < -S_{x}^{11}$  for $T <  0.02$.
 The interband contribution's 
 $S_{\nu}^{12} (= S_{\nu}^{21})$
 is    much smaller than those of the intraband contribution.

\begin{figure}
  \centering
\includegraphics[width=8cm]{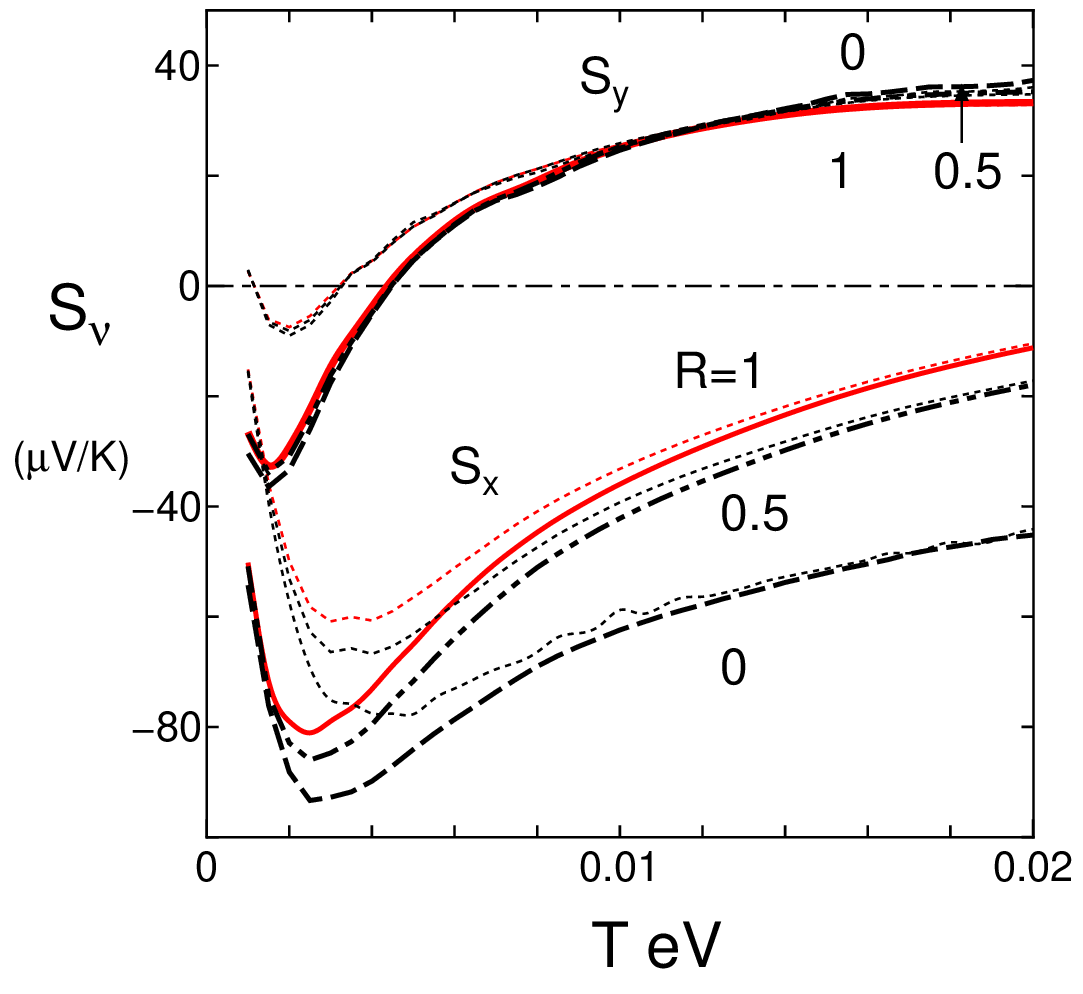} 
     \caption{(Color online )
$T$  dependence of the Seebeck coefficients $S_x$ and $S_y$, 
 which is calculated using  Eqs.~(\ref{eq:S}) and (\ref{eq:S_comp})
 for $R$ = 1 (solid line), $R$ = 0.5 (double-dot-dashed line) and $R$ = 0 
 (dashed line).  
 The chemical potential $\mu$ is estimated from 
 the inset of Fig.~\ref{fig2}.
$S_y$ exhibis  a sign change, while $S_x < 0$ for $T < 0.02$.
The dotted line is obtained for 
 transfer energy with only Re $w_h$. 
}
\label{fig5}
\end{figure}

Next, we examine 
 the $T$  dependence of the Seebeck coefficients $S_x$ and $S_y$
 in Eq.~(\ref{eq:S}). 
 Figure \ref{fig5} shows 
  $S_\nu$,  
  where solid line, double-dot-dashed line and dashed line 
  correspond to   
 $R$ = 1, 0.5, and 0, respectively. 
 These lines  are calculated  for 
 transfer energies  with both Re $w_h$ and Im $w_h$ in Table \ref{table_1}, 
 while the dotted line is calculated  for 
 transfer energies  with only Re $w_h$.
 It turns out that $S_y > S_x$ at any temperature. 
 Noticeable  behaviors are  as follows. 

(1)
 At low temperatures, 
  both $S_y (<0)$ and $S_x (<0)$  decrease from zero and take a minimum. 
 With a further increase of $T$, both $S_y$ and $S_x$ start increasing 
   and become close to their respective dotted lines.   

(2) At high temperatures, 
 $S_y$  shows  a sign change with increasing $T$. 
  For $R$ = 1,  
 $S_y$ undergoes  a sign change  at $T \simeq 0.004$, where 
 $S_y < 0$  for $T < 0.004$ and  
 $S_y > 0$  for $T > 0.004$.
 This result of  $S_y > 0$ at high temperatures 
 is also seen in Fig.~\ref{fig4}, 
 where  $S_{y}^{22}(T) > - S_{y}^{11}(T)$ for $T > 0.004$. 
 The difference in  $S_y(T)$ between 
 $R$ = 1 and 0 is negligibly small, where 
 $R$ = 1 denotes weak e--p coupling 
and  $R$ = 0  corresponds to the absence of the e-p coupling. 
Thus, the $T$ dependence of $R$ = 1    is determined 
   by the Fermi distribution function, $f(\ep - \mu)$, 
   and 
the effect of the e-p  interaction on $S_y$  
is negligibly small.
On the other hand,  
 there is no sign change for $S_x$, where  
   $S_x < 0$ in the region of $T < 0.02$. 
 Further, the decrease in $R$ leads to a noticeable  reduction in $S_x(T)$. 
  The case of $R$ = 0.5 suggests 
    continuous variation of $S_\nu$ 
    with decreasing $R$.

(3) The role of the SOC is discussed. 
 The solid, double-dot-dashed  and dashed 
 lines are obtained for the transfer energy 
 with both Re $w_h$ and Im $w_h$, 
 while  dotted  lines are  obtained for 
 transfer energy with only Re $w_h$.
Noting  the former lines are  slightly lower  than the latter lines, 
   we can see the SOC enhances the absolute value 
 of the Seebeck coefficient (for $S_\nu < 0$). 
  This effect of the SOC  becomes large 
 at low temperatures. 

In the following we discuss only the case with $R = 1$, 
 which includes the e-p interaction, 
because both  $L_{11}^x(T)$ and  $L_{11}^y(T)$ 
 with  $R = 0$ increase monotonously with increasing $T$,~\cite{Suzumura_Tsumuraya_2021}
 which is incompatible with the  experiment.~\cite{Inokuchi1995}
 
\begin{figure}
  \centering
\includegraphics[width=5cm]{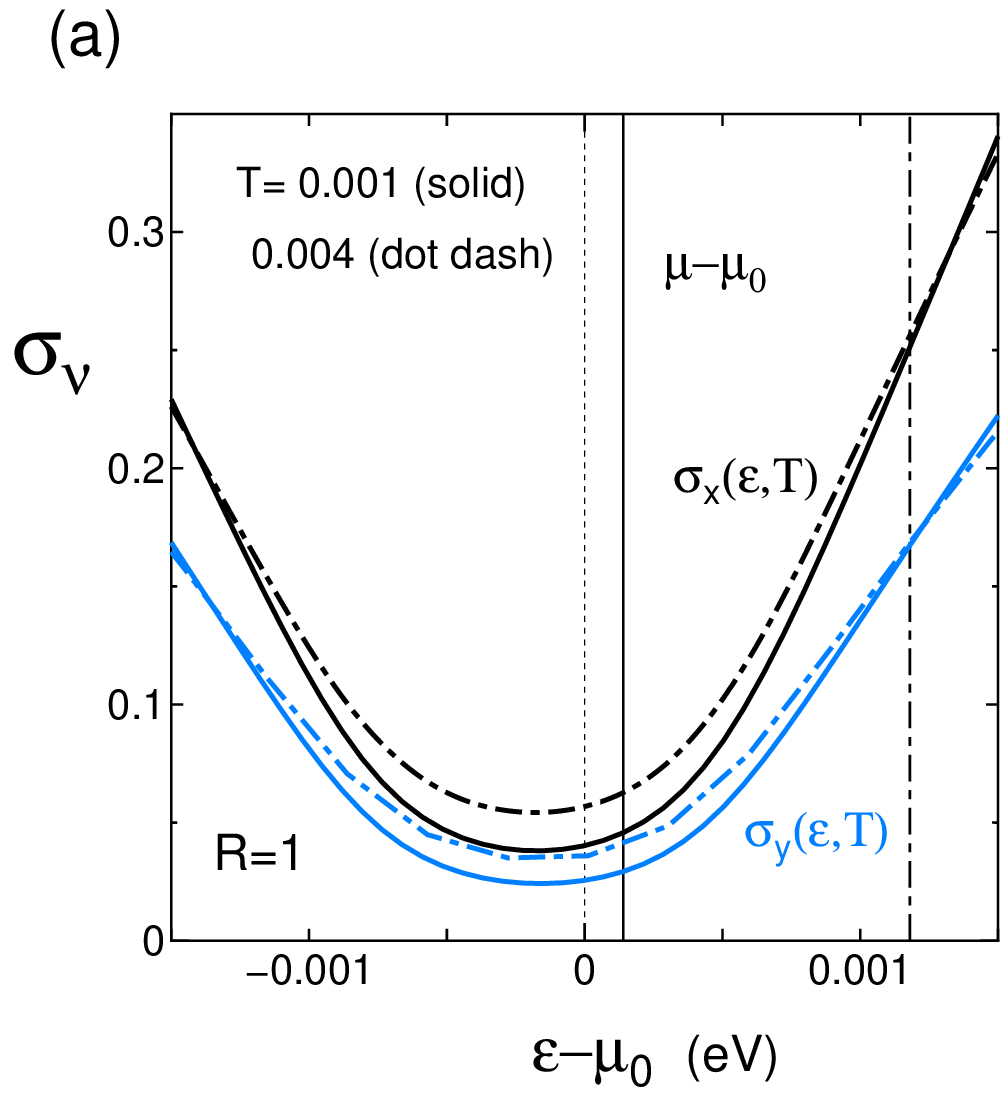}\\
\includegraphics[width=5cm]{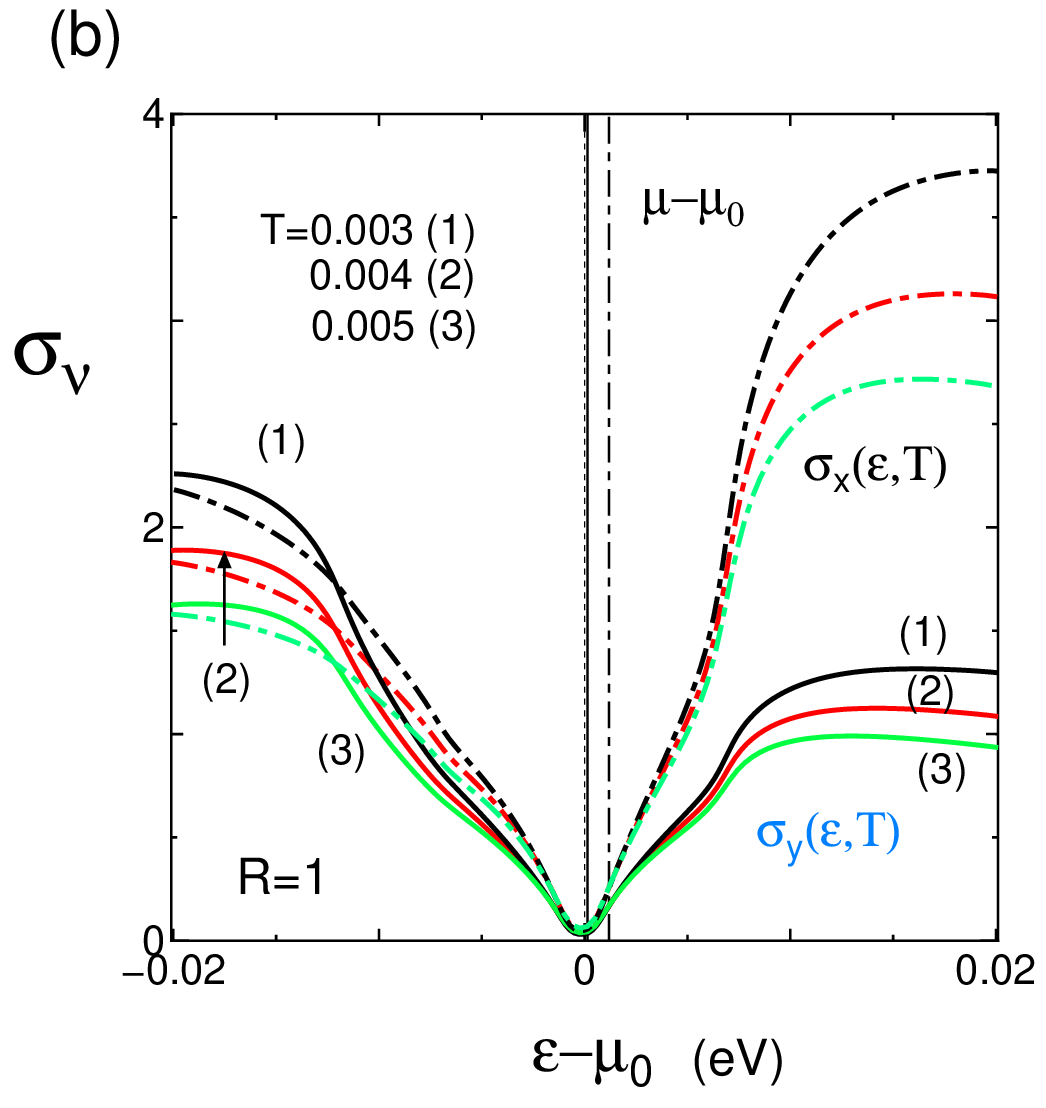}\\
\includegraphics[width=5cm]{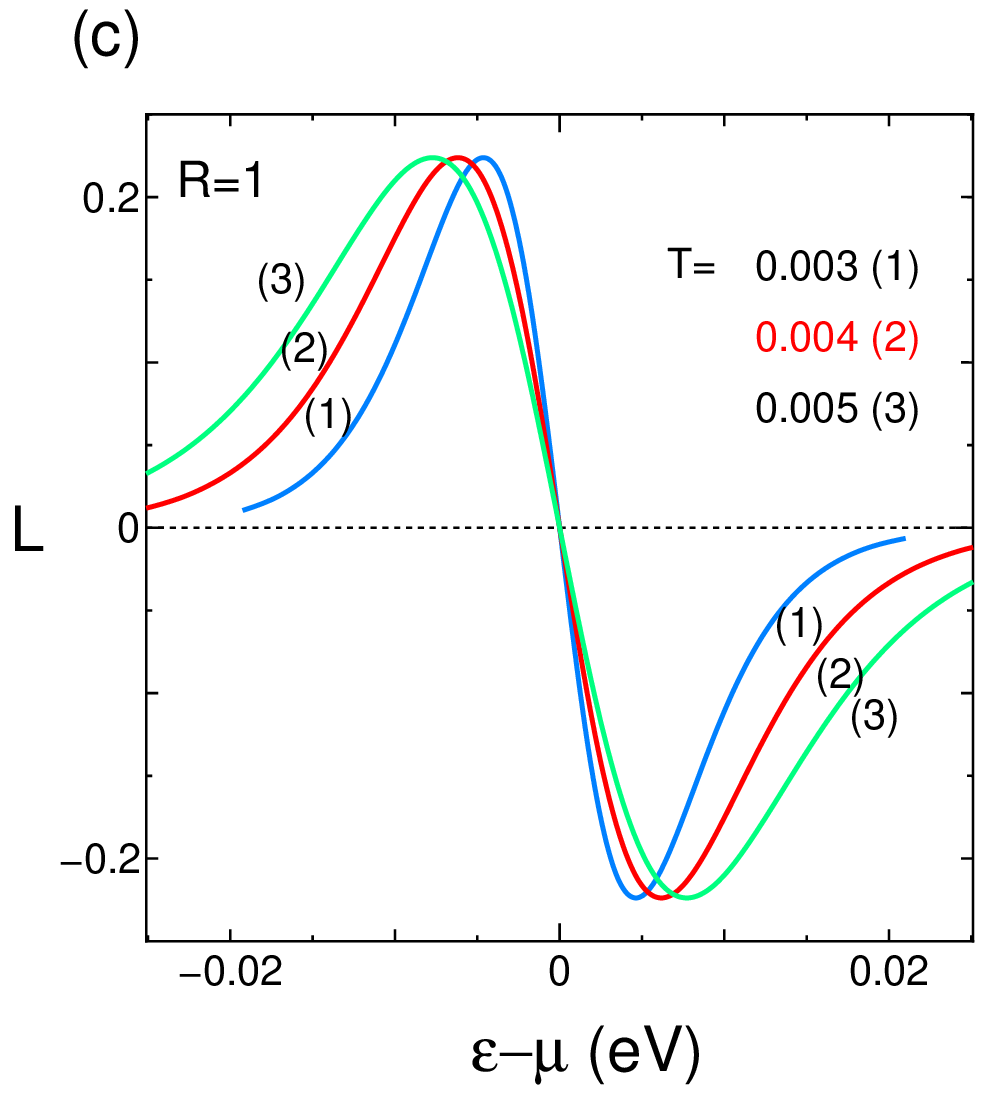}
     \caption{(Color online)
(a) Spectral conductivity $\sigma_\nu(\ep,T)$ ($\nu$ = $x$ and $y$)  
   as a function of $\ep - \muz(=\tep)$ with $|\tep| < 0.0015$
for $T$ = 0.001 and 0.004.  
(b) Spectral conductivities $\sigma_y(\ep,T)$ (solid line) 
 and $\sigma_x(\ep,T)$ (dot-dashed line)  
   as a function of $\ep - \muz(=\tep)$ with $|\tep| < 0.02$ for 
  (1) $T$ = 0.003, (2) 0.004,  and (3) 0.005.  
The following  asymmetry with respect to $\tep$:  
 By taking   $\tep > 0$, one finds 
  $\sigma_x(\tep+\muz,T) \gg  \sigma_y(\tep+\muz,T)$  
  and  
 $\sigma_x(\tep+\muz,T) > \sigma_x(-\tep+\muz,T) 
\simeq  \sigma_y(-\tep+\muz,T)>\sigma_y(\tep+\muz,T)$. 
(c) $\ep - \mu$ dependence of 
$L$  [Eq.~(\ref{eq:Factor})]
 for fixed (1) $T$ = 0.003, 
 (2) 0.004, and (3) 0.005, corresponding to (b). 
}
\label{fig6}
\end{figure}

\section{Analysis of the Seebeck coefficients}

To understand the characteristic behavior of the Seebeck coefficient in Fig.~\ref{fig5}, which is 
different from the Seebeck coefficient in \ET, 
we study the spectral conductivity.
 Noting 
  the main contribution 
 among the  components of the Seebeck coefficient 
 in Fig.~\ref{fig4},
 comes from intraband process (i.e.,$\gamma=\gamma'$),  
 we rewrite  
$ v_{\gamma \nu}(\bk) \equiv v^{\nu}_{\gamma \gamma}(\bk)$ 
 and write the  spectral conductivity,  $\sigma_\nu(\ep,T)$, as  
\begin{eqnarray}
   \sigma_\nu(\ep,T)  &=&   
     \frac{e^2 }{\pi \hbar}  \sum_{\gamma =1}^2 \frac{1}{N} \sum_{\bk} 
         \sigma_{\g\nu}(\bk,\ep,T)\; ,    \label{eq:22m} 
\end{eqnarray}
 where 
\begin{eqnarray}
    \sigma_{\g\nu}(\bk,\ep,T) &=&  v_{\gamma \nu}(\bk)^2 
        \left( \frac{\Gamma_\g}{(\ep - E_{\gamma}(\bk))^2 + \Gamma_\g^2} \right)^2     \; .   \label{eq:23a}
        \nonumber \\
\end{eqnarray}

 The $T$ dependence of the spectral conductivity $\sigma_\nu(\ep,T)$ 
 comes from  $\Gamma_{\rm ph}^\g$
 of Eq.~(\ref{eq:eq16a}) and  $\Gamma_\g = \Gamma + \Gamma_{\rm ph}^\g$.
Note  the special case, where 
  the conduction and valence bands have   symmetry
     given by 
       $E_{1}(\bk_1)-\muz = -[E_{2}(\bk_2)-\muz]$ 
       and $v_{1\nu}(\bk_1) = v_{2\nu}(\bk_2)$
 with $\bk_1-\bkD=-\bk_2+\bkD.$ In this symmetric case,  one obtains  
 $\sigma_\nu(\tep+\muz,T)= \sigma_\nu(-\tep+\muz,T)$ 
    with $\tep =\ep-\muz$ leading to  $L_{12}^\nu  = 0$ from 
 Eq.~(\ref{eq:L12}).
 Although such symmetry  is  found in 
   a Dirac cone with a linear spectrum, 
    the present model has band asymmetry, which  results in 
   $\sigma_\nu(\tep+\muz,T) \not= \sigma_\nu(-\tep+\muz,T)$ 
 and the deviation of the minimum of $\sigma_\nu(\ep,T)$ from  $\ep=\muz$.   

 First, we examine  $S_\nu(T)$ at low temperatures, where   
   $S_x<0$ and   $S_y<0$  in Fig.~\ref{fig5}, which is in sharp contrast to  
   those in \ET, which have $S_x>0$ and   $S_y>0$.
  Figure  \ref{fig6}(a) shows  
    the spectral conductivity of $\sigma_x(\ep,T)$  and $\sigma_y(\ep,T)$ 
  for small $|\ep - \muz|$   with  fixed $T$ = 0.001 and 0.004.
 They show a common feature in which  a minimum 
   of $\sigma_\nu(\ep,T)$ ($\nu = x$ and $y$) is located  at 
   $\ep$ being slightly lower than $\muz$, which is ascribed 
    to the deviation from the linear band  around the Dirac point. 
  The  inequality  of  $\sigma_x(\ep,T)>\sigma_y(\ep,T)$ is 
consistent with $v_x>v_y$ as shown 
 in Fig.~\ref{fig3}(c).
 As in the case of \ET~\cite{Suzumura_JPSJ_2024}, we discuss 
 the Seebeck coefficient at low temperatures using the Mott formula~\cite{Mahan1980}, 
 $S_\nu = - (\pi^2/3e) T \sigma'_{\nu}(\mu,T)/\sigma_\nu(\mu,T)$.
 As shown in the inset of Fig.~\ref{fig2}, $\mu > \muz$.
 Therefore,   we find  that $S_\nu(T) < 0$  due to   $\sigma'_{\nu}(\mu,T) > 0$ 
  in Fig.~\ref{fig6}(a), with  $\ep=\mu(T)$ (the vertical line).  
  This indicates that the Seebeck coefficient is 
``electron-like''. 
  On the other hand, in \ET, $\mu < \muz$. As a result, 
  the Seebeck coefficient is ``hole-like'', or $S_\nu(T) > 0$.
  Furthermore, the origin of $\mu>\muz$ in \BETS\ is the 
  presence of the shoulder $C$ in the DOS as shown in Fig.~\ref{fig2}, 
  which originates from the presence of the peak at $Y$ point just below 
 $\muz$. 

Next, we examine $S_\nu(T)$ at higher temperatures.  
  Figure \ref{fig6}(b) shows  
   $\sigma_\nu(\ep,T)$ with large $|\ep-\muz|$, 
 which provides $L_{\rm 12}^\nu$ at higher temperatures.
The important feature in Fig.~\ref{fig6}(b) is that 
  $\sigma_x(\ep,T)$ is much larger than  $\sigma_y(\ep,T)$ for $\ep-\muz>0$.
  Furthermore, $\sigma_y(\ep,T)$ for $\ep-\muz>0$ is smaller than 
  $\sigma_y(\ep,T)$ for $\ep-\muz<0$.
  The shoulder near $\ep-\muz =0.007$ corresponds to the energy of the 
  Van Hove singularity at $A$ in Fig.~\ref{fig2}. 
  We consider these features in $\sigma_\nu(\ep,T)$ to be the 
  physical origin of the sign change of $S_y$, as discussed below. 
  
 As seen from  Eq.~(\ref{eq:L12}), $L_{12}^\nu$ is calculated from 
   $\sigma_\nu(\ep,T)$ multiplied by
   \begin{equation}
 L =  \left( \frac{\partial f(\ep-\mu) }{\partial \ep} \right) (\ep-\mu) 
  \; , 
 \label{eq:Factor} 
\end{equation}
which is shown in Fig.~\ref{fig6}(c) at (1) $T=0.003$, (2) $0.004$, 
and (3) $0.005$.
As temperature increases, the peak position of Eq.~(\ref{eq:Factor}) moves to 
 larger $|\ep - \mu|$. As a result, the center of the weight for $\sigma_y(\ep,T)$ 
also moves to  larger $|\ep - \mu|$. 
We can see that $L_{12}^x$ is negative for every temperature 
 after the integral 
with respect to $\ep$. On the other hand, $L_{12}^y$ changes its sign at $T=0.004$. 
Actually,  for $T$ = 0.003, the electron contribution ($\ep-\mu>0$) is larger than 
the hole  contribution ($\ep-\mu<0$),
while for $T$ = 0.005, the electron  contribution  is than the hole  contribution.

We examine  the difference 
 between  $\sigma_y(\ep,T)$  and  $\sigma_x(\ep,T)$ in the region of 
 $\ep-\muz>0.007$ in Fig.~\ref{fig6}(b) 
  in terms of 
 the velocity 
${\bf v}_\g(\bk)$ ($\g$= 1 and 2),  
 which is directly  related to 
  $\sigma_\nu(\ep,T)$ by  Eq.~(\ref{eq:23a}).  
 Note that the velocity is calculated from  
\begin{eqnarray}
 {\bf v}_\g(\bk)= (v_{\g x},v_{\g y})
=  \left( \frac{ \partial E_\g(\bk)}{
                                         \partial k_x},
      \frac{\partial E_\g(\bk)}{\partial k_y}  \right)
                 \; .
  \label{v12}
\end{eqnarray}
 Since  ${\bf v}_\g$ is perpendicular to 
 the tangent of the contour line of $E_\g (\bk)$, 
 the ratio $v_{\g x}/v_{\g y}$ is equal to  the slope of the tangent  line. 
 At $\ep - \muz \sim 0.007$, the dominant contribution 
 comes from the DOS with  a Van Hove singularity at $A$ in Fig.~\ref{fig2}.
 From Fig.~\ref{fig3}(d), the ratio of $v_{1y}/v_{1x}$ near the $M'$ point 
 is given by  $\sim 0.54$, leading  to 
$(v_{1x}/v_{1y})^2 \simeq 3.5$, which 
 is larger than $\sigma_x(\ep,T)/\sigma_y(\ep,T) \simeq 2.3$  
  for $\ep- \muz   \simeq 0.0074$. 
 This difference is reduced if 
 the former is averaged on the contour 
 with $E_1- \muz \simeq 0.0074$ in Fig.~\ref{fig3}(d).  
 On the other hand, 
 the fact that $\sigma_y(\ep,T) \simeq \sigma_x(\ep,T)$ 
  for  $\ep - \muz \simeq - 0.014$ is understood as follows. 
 In this case, the dominant contribution  comes from 
 the DOS with  a Van Hove singulaity at $B$ in Fig.~\ref{fig2}.
 From Fig.~\ref{fig3}(e), the ratio  $v_{1y}/v_{1x}$ near the $Y$ point 
  and two saddle points 
  is given by  $\simeq 1.0$ leading  to 
$(v_{1x}/v_{1y})^2 \sim 1.0$, which 
 is almost equal to $\sigma_y(\ep,T)/\sigma_x(\ep,T) \simeq 1$ 
  for $\ep -\muz  \simeq - 0.014$. 

We conclude the present calculation as  follows. 
 The results $S_y > 0$ and $S_x < 0$  for \BETS \; at high temperatures 
 are understood from the anisotropy of the spectral conductivity,
 $\sigma_\nu(\ep,T)$ with respect to $\ep$, 
  which comes from the difference in  
     the ratio  $v_x/ v_y$ between  the conduction   and valence bands. 
The conduction band gives  $|v_x/ v_y| \gg 1$ 
 due to  a large inclination of the  line connecting the two Dirac points 
 and the $M'$ point of the TRIM, while 
 the valence band gives  $|v_x/ v_y| \simeq  1$ 
 due to a moderate inclination  $\simeq 1$ 
 of  the line connecting the two Dirac points, two saddle points, 
 and the $Y$ point of the TRIM.

\section{Summary and Discussion}
We examined the $T$ dependence of the Seebeck coefficient 
for the Dirac electrons in \BETS \; at ambient pressure using a TB model obtained  
from a first-principles DFT calculation. 
 A summary and discussion are  as follows. 

 At low temperatures, the sign of the Seebeck coefficient is negative 
 in both the $x$ and $y$ directions, 
i.e., $S_x<0$  and $S_y<0$, indicating the electron  behavior.  
  The sign of the Seebeck coefficient is determined 
 not only by the sign  of $\mu- \muz$ 
 but also by the  velocities  $v_{\g \nu}(\bk)$ 
 of the conduction and valence bands. 
 Thus,   we calculated  the spectral conductivity, $\sigma_\nu(\ep,T)$  at low temperatures and found  that  
 the energy derivative of $\sigma_\nu(\ep,T)$ with respect to $\ep$ 
 at $\ep = \mu$  takes a positive sign,  leading to   $S_\nu < 0$. 
 This comes from  the fact that the chemical potential $\mu(T)$ becomes 
 larger than $\muz$ because of the presence of the shoulder 
   just below $\muz$ in the DOS 
   (i.e., the energy maximum  at $Y$ point.) 
 This is in sharp contrast to the case of \ET, in which $\mu(T)$ becomes smaller  than $\muz$.

 At high temperatures,  we obtained $S_y >0$ and $S_x < 0$, where 
  $S_y$ undergoes a sign change. 
 This  difference is understood  from  the spectral conductivity 
 with large $|\ep - \mu|$ showing that  
  $\sigma_y(\ep,T)$ with $\ep > \mu$  
 is much smaller 
than 
  $\sigma_y(\ep,T)$ with $\ep < \mu$.  
Furthermore,  $S_\nu$ was analyzed 
in terms of the spectral conductivity and the factor $\left( 
\frac{\partial f}{\partial \ep} \right) (\ep-\mu)$. 
 $S_y=0$ at $T$ = 0.004 is interpreted as  
 the contribution from the valence band becoming 
   equal to  that of the  conduction   band,  
 while, for $S_x$, 
  the contribution from the valence band is always smaller than 
     that of the conduction band.
 The present result was obtained  
 for  weak e--p coupling ($R \sim 1 $), 
 which is  a reasonable choice 
 to reproduce 
  the  conductivity at high temperatures~\cite{Suzumura_Tsumuraya_2021} 
    in  an organic conductor.    
 
Finally,  we compare the  present work on BETS  with 
 that by Ohki {\it et al.}~\cite{Kobayashi_2020}.  
They examined the electronic states at low temperatures, where   
  the  transfer energies are derived in the absence of  SOI and  
 the SOI is  taken into account as a tunable parameter. 
 The on-site repulsive interaction  is introduced 
  to examine the spin-ordered state.
 In the present work, 
  we examined  the electronic state mainly  at high temperatures, 
 where the on-site repulsive interaction is discarded  
 due to the normal state.    
 In terms of the fully relativistic treatment of  
 the first-principles density functional calculation, 
 the SOC is obtained without adjustable parameters. 
  Further, the e-p scattering is  taken into account,   
  which increases with temperature.

\acknowledgements
We thank T. Furukawa and N. Nishikawa (Tohoku University)
  for useful discussions and sending us data for the Seebeck coefficient  
 of BETS.  
Y.S. is indebted to   H. Fukuyama for  valuable comments on the spectral conductivity.
This work was supported
by a Grant-in-Aid for Scientific Research 
(Grants No. JP23H01118 and No. JP23K03274)
and JST-Mirai Program Grant No. JPMJMI19A1.

\appendix

\section{Matrix Elements  in the TB Model}
  With  first-principles calculations, 
 the TB model is obtained as~\cite{Tsumuraya_Suzumura}
\begin{eqnarray}
H_0 &=& \sum_{i,j = 1}^N \sum_{\alpha, \beta } \sum_{s, s'= \pm}
 t_{ij; \alpha s,\beta s'} a^{\dagger}_{i \alpha s} a_{j \beta s'}    
                           \nonumber \\
 &=&
  \sum_{\bk} \sum_{s, s'} \sum_{\alpha, \beta}  
      h_{\alpha \beta; s s'}(\bk)  a^{\dagger}_{\alpha s}(\bk) 
         a_{\beta s'}(\bk)\; . 
\label{eq:A1}
\end{eqnarray}

 We introduce  site-potentials acting on the $B$ and $C$ sites, $\Delta V_{B}$ and $\Delta V_{C}$, which are measured from site-energy at $A$ and $A'$ site, $V_{A}$: 
\begin{eqnarray}
\Delta V_{B} = V_{B} - V_{A}, \\
\label{siteVb}
\Delta V_{C} = V_{C} - V_{A},
\label{siteVc}
\end{eqnarray}
where $V_{A}$, $V_{B}$, and $V_{C}$ are the site-energies at each molecule, 
 which   are calculated using MLWFs $|\phi_{\alpha,0} \rangle$; 
\begin{eqnarray}
V_{\alpha} =\langle\phi_{\alpha, \sigma,0}| H |\phi_{\alpha, \sigma^{\prime},0} \rangle, 
\label{eq:eqV33}
\end{eqnarray}
where ${\alpha}$ = $A$ (=~$A^{\prime}$), $B$, and $C$.  molecules. 
These transfer energies  are  listed in Table~\ref{table_1}, where 
$\Delta V_{C}$  is modified from 0.0208 due to 
 a correlation effect~\cite{Tsumuraya_Suzumura}.   
 
 Since the case of SOC  with only the same spin is  examined, 
we rewrite Eq.~(\ref{eq:A1}) as (discarding spin)
\begin{eqnarray}
 H_0 = \sum_{\bk}\sum_{m=1}^4   \sum_{n=1}^4
  h_{mn}(\bk) a_m^\dagger(\bk) a_n(\bk)
\; ,
\end{eqnarray}
 where 
   $a_{\alpha s}$ is replaced by  $a_{m}$  and  
 $m, n$ =1, 2, 3, and 4 correspond to 
  $A$, $A'$, $B$ and $C$, respectively . 

With transfer energies given in Table \ref{table_1},  
 a matrix element $h_{mn}(\bk)$   is  calculated from 
   Eq.~(\ref{eq:def_transfer}), where 
  nearest-neighbor and 
next-nearest-neighbor sites (Fig.~\ref{fig1}) 
 are taken into account~\cite{Tsumuraya_Suzumura}.   
 The  $y$ axis is taken as the negative direction  to make 
  the result  consistent 
 with the previous case~\cite{Tsumuraya_Suzumura,Suzumura_Tsumuraya_2021}.  

 In terms of  
    $X= \e^{i\kx}$, $\bar{X}=  \e^{-i\kx}$, 
    $Y= \e^{i\ky}$, and $\bar{Y}=  \e^{-i\ky}$, 
  matrix elements $h_{mn}$ are given by 
 \begin{eqnarray}
 h_{11} & = & h_{22} =
  a_{1d}(Y+\bar{Y})
 +s1^*X+s1\bar{X} 
                          \; , \nonumber \\
 h_{33}&=&   a_{3d}(Y+\bar{Y})
     + s3(X + \bar{X})+ \Delta V_B
                           \; , \nonumber \\
 h_{44}&=&    a_{4d}(Y+\bar{Y})
  + s4(X + \bar{X})+ \Delta V_C 
                           \; , \nonumber \\
 h_{12}&=&  a_3+a_2Y+d_0\bar{X}+ d_1XY 
                          \; , \nonumber \\
h_{13}&=&    b_3+b_2\bar{X}+ c_2 \bar{X} Y + c_4\bar{X}\bar{Y}
                           \; , \nonumber \\
 h_{14}&=&   b_4Y+b_1\bar{X}Y+c_1\bar{X} + c_3
                           \; , \nonumber \\
h_{23} &=&   b_2+b_3\bar{X}+c_2 \bar{Y} + c_4 {Y}
                           \; , \nonumber \\
h_{24} &=&   b_1+b_4\bar{X}+c_1Y + c_3\bar{X}Y
                           \; , \nonumber \\
h_{34}&=&   a_1+a_1Y + d_2\bar{X}+d_3X 
                   + d_2 XY+d_3\bar{X}Y 
                               \; ,   \nonumber \\
\end{eqnarray}
 and 
 $h_{ji} = h_{ij}^*$.



\end{document}